\documentclass[fleqn,usenatbib]{mnras}

\usepackage{newtxtext,newtxmath}
\usepackage[T1]{fontenc}

\usepackage{graphicx}	% Including figure files
\usepackage{amsmath}	% Advanced maths commands
\usepackage{subcaption}
\usepackage{caption}
\usepackage{xcolor}
\definecolor{red}{RGB}{227,37,107}%razzmatazz
\definecolor{peri}{RGB}{10,11,216}
\definecolor{smar}{RGB}{114, 137, 218}

\title[The 21\,cm Signal in a Changing Environment]{Global 21\;cm Signal Recovery Under Changing Environmental Conditions}

\author[J. H. N. Pattison et al.]{
Joe H. N. Pattison$^{1,2}$\thanks{E-mail: jhnp2@cam.ac.uk},
Jean Cavillot $^{3}$,\thanks{E-mail: jean.cavillot@uclouvain.be}
Harry T.J. Bevins$^{1,2}$\thanks{E-mail: htjb2@cam.ac.uk},
Dominic J. Anstey$^{1,2}$\thanks{E-mail: da401@cam.ac.uk},
Eloy de Lera Acedo$^{1,2}$\thanks{E-mail: ed330@cam.ac.uk}
\\
$^{1}$Astrophysics Group, Cavendish Laboratory, J.J. Thomson Avenue, Cambridge, CB3 0HE, UK\\
$^{2}$Kavli Institue for Cosmology, Madingley Road, Cambridge, CB3 0HA, UK\\
$^{3}$Antenna Group, Université Catholique de Louvain, Louvain-la-Neuve, Belgium
}

\date{Accepted XXX. Received YYY; in original form ZZZ}

\pubyear{2024}

\begin{document}
\label{firstpage}
\pagerange{\pageref{firstpage}--\pageref{lastpage}}
\maketitle

% Abstract of the paper
\begin{abstract}
The redshifted 21\;cm line of cosmic atomic hydrogen is one of the most auspicious tools in deciphering the early Universe.
Recovering this signal remains an ongoing problem for cosmologists in the field, with the signal being hidden behind foregrounds approximately five orders of magnitude brighter than itself.
A traditional forward modelling data analysis pipeline using Bayesian data analysis and a physically motivated foreground model to find this signal shows great promise in the case of unchanging environmental conditions.
However we demonstrate in this paper that in the presence of a soil with changing dielectric properties under the antenna over time, or a changing soil temperature in the far field of our observation these traditional methods struggle.

In this paper we detail a tool using Masked Auto-regressive Flows that improves upon previous physically motivated foreground models when one is trying to recover this signal in the presence of changing environmental conditions.
We demonstrate that with these changing parameters our tool consistently recovers the signal with a much greater Bayesian evidence than the traditional data analysis pipeline, decreasing the root mean square error in the recovery of the injected signal by up to 45\%.
\end{abstract}

% Select between one and six entries from the list of approved keywords.
% Don't make up new ones.

%%%%%%%%%%%%%%%%%%%%%%%%%%%%%%%%%%%%%%%%%%%%%%%%%%

%%%%%%%%%%%%%%%%% BODY OF PAPER %%%%%%%%%%%%%%%%%%

\section{Introduction}
Between the emission of the cosmic microwave background and the emergence of the first galaxies the 21 cm signal from cosmic hydrogen is expected to be one of the most useful tools in understanding the early Universe.
The detection of this signal, however, is an ongoing challenge facing cosmologists.
While a complete tomographic map of cosmic hydrogen between cosmic dawn and reionisation remains a distant dream to astrophysicists, the detection of the sky-averaged monopole signal is one closer at hand.
A description of the depth, width and general shape of this globally averaged signal would go to unlocking information about early galactic structure formation \citep{Yajima2014DistinctiveQuasars}, dark matter \citep{Barkana2018Signs21-cm}, and Population III stars \citep{Gessey-Jones2022ImpactSignal}.

A potential detection of this signal was made by the Experiment to Detect the Global Epoch of reionisation Signal \citep[EDGES,][]{Bowman2018AnSpectrum}, though this detection was in tension with the standard cosmological model -  showing a very different shape and depth to our standard theoretical models \citep[e.g.][]{Cohen2017ChartingSignal}.
Since this detection a number of alternate cosmological models have been proposed to resolve the tensions between the standard model of cosmic evolution and the observed signal, including but not limited to weakly interacting dark matter \citep[e.g.][]{Liu2019RevivingCosmology}, enhanced radio backgrounds \citep{Fialkov2019SignatureSpectrum, Mittal2022ImplicationsSurveys}, and competing sources of reheating in the early Universe \citep{Gessey-Jones2023SignaturesObservables}.

As of yet no follow up experiments have been able to confirm this detection; with \citet{Singh2017SARASSignal} rejecting the EGDES detection with a 95\% confidence.
In no small part this is the result of the overwhelming foregrounds that obscure the signal from cosmic atomic hydrogen.
Galactic foregrounds, predominantly from the synchrotron emission of electrons caught in strong magnetic fields, overwhelms the 21 cm signal by approximately five orders of magnitude, meaning a precise and accurate description of these foregrounds is needed to resolve the signal.
This foreground modelling is complicated by a number of factors, including antenna chromaticity \citep{Anstey2020AExperiments, Cumner2023TheExperiment}, ionospheric effects \citep[e.g.][]{Shen2021QuantifyingObservations}, man-made radio-frequency interference (RFI) \citep{Leeney2022AMitigation, Anstey2023EnhancedReweightingb}, and a physical horizon \citep{Bassett2021LostAnalysis, Pattison2024ModellingOunds}.

All of these sources of signal obfuscation have significant time varying properties which must be corrected for: dielectric properties of the soil in the near field affecting the structure of the beam \citep{Spinelli2022AntennaReconstruction} which change with Terran weather, solar UV and x-ray emission altering the properties of the ionosphere which will vary with solar weather \citep{Liu2011SolarReview}, transient RFI sources including plane and satellite radio communication which must be identified and removed \citep{Anstey2023EnhancedReweightingb}, and changing soil temperatures and dielectric properties in the far field overnight or as seasons progress.

This work expands on \citet{Anstey2023UseModelling} and \citet{Pattison2024ModellingOunds} aiming to account for the time varying properties of the horizon and antenna beam given the use of a physically motivated foreground model.

Section \ref{sec:methods} describes how we generate the beams used to describe the antenna in our mock data, discusses how a traditional physically motivated Bayesian data analysis pipeline would function, and discusses the implementation of normalising flows into a novel analysis pipeline to allow signal recovery with changing environmental conditions.
Section \ref{sec:results} goes through our results, discussing how the two pipelines recover a 21\;cm signal with changing soil moisture levels and temperatures over a series of observations.
Section \ref{sec:conclusions} outlines our key conclusions and discusses where we believe this work will lead in future.

\section{Methods}
\label{sec:methods}
This section details the methods used in this paper to analyse mock data generated in a physically motivated Bayesian data analysis pipeline. 
The beams used to in this paper are made following the methodology in Section \ref{sec:beams}, and the data simulation process using these beams is discussed in Section \ref{sec:simulation}.
Sections \ref{sec:bayesian} and \ref{sec:LST} discuss Bayesian statistics and the physically motivated Bayesian forward modelling pipeline used for REACH.
Sections \ref{sec:flows} and \ref{sec:novel} discuss normalising flows and their application to expand upon the Bayesian analysis pipeline.

\subsection{Beam Generation}
\label{sec:beams}
The beams used in this paper have been obtained after solving the Method of Moments (MoM) \citep{Harrington1993FieldMethods}. 
In this full-wave solver, the antenna is described using a set of elementary basis functions \(J_i\). 
Using Galerkin testing, we calculate a matrix of impedance \(Z_\text{MoM}\) which represents the interactions between basis functions and testing functions. 
Here we assume the presence of a semi-infinite soil with a given permittivity below the antenna. 
Thus, a spectral formulation, provided in \citep{Cavillot2019FastPlane}, is adopted where the effect of the soil is accounted for in the calculation of the entries of \(\bf{Z_\text{MoM}}\) thanks to reflection coefficients.

In calculating the total current over the antenna \(J\), where \(J\) is made by summing the basis functions multiplied by their corresponding coefficients, \(c_i\),

\begin{equation}
    J = \sum_ic_iJ_i,
\end{equation}

\noindent we must invert the impedance matrix and multiply by the excitation vector to yield our coefficients: \(\bf{c}= \bf{Z_\text{MoM}^{-1}v}\).
In this case, \(\bf{v}\) is obtained by applying a delta-gap excitation on the antenna. 
Once the vector of coefficients \(\bf{c}\) is obtained, the beams can be generated by performing the Fourier Transform of the current distribution and by multiplying it by the Green's function of the layered medium \citep{Cavillot2020EfficientPlane}

\begin{figure}
    \centering
    \includegraphics[width = \linewidth]{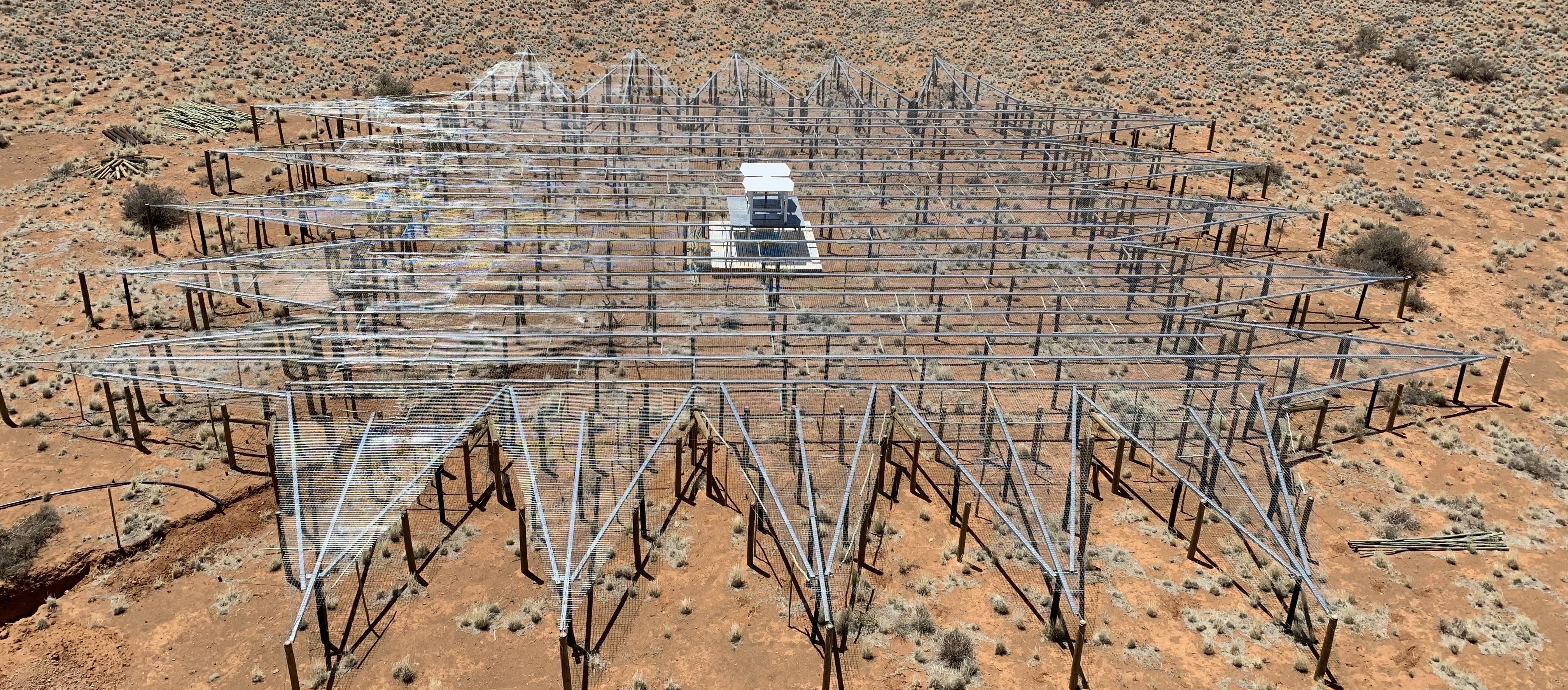}
    \caption{Overhead image of the REACH hexagonal dipole antenna over its ground plane, situated in the Karoo radio reserve in South Africa. Image courtesy of Dr. Saurabh Pegwal.}
    \label{fig:reach}
\end{figure}

\begin{figure}
    \centering
    \includegraphics[width = \linewidth]{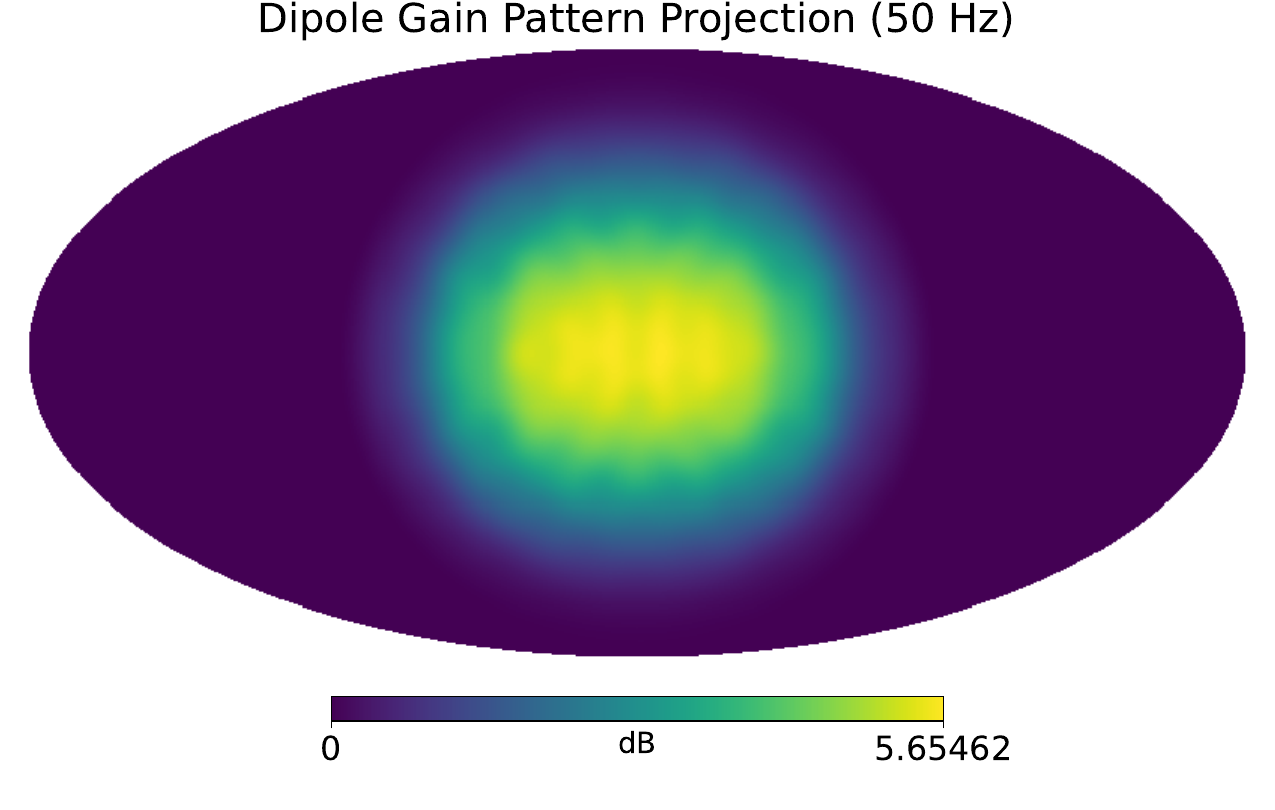}
    \caption{Projection of the antenna gain pattern at 50\,MHz of the REACH hexagonal bladed dipole antenna with a ground plane 1\,m above a semi infinite layer of soil with a dielectric constant of \(\epsilon = 3.8 + 0.29j\).}
    \label{fig:beampattern}
\end{figure}

In this case we build a beam analogous to the REACH dipole, which we see in as shown in Figure \ref{fig:reach}.
It is modelled as hexagonal bladed dipole antenna; each blade with dimension of 1400\,mm by 928\,mm on a 20 x 20\,m metallic ground plane with serrated edges, 1\,m over a semi-infinite layer of soil \citep{Cumner2022RadioCase}.
We show an example of the beam pattern produced in Figure \ref{fig:beampattern}.

\subsection{Data Simulation}
\label{sec:simulation}

Simulation of the data we analyse in this paper follows \citep{Pattison2024ModellingOunds}.
We generate a map of the variation of the spectral indices over the sky using the 2008 Global Sky Model \citep{deOliveira-Costa2008AGHz}.
We do this by mapping the 408\,MHz onto a corresponding map at 230\,MHz to make sure our map is uncontaminated by the cosmological 21\,cm signal.

From this map we subtract any power originating from the cosmic microwave background (CMB), and rotate the map according to the time, date and location of observation.
This map is then masked by the horizon surrounding the antenna, generated using the \textsc{shapes} algorithm \citep{Bassett2021LostAnalysis}, which is given a power equivalent to the expected soil temperature.
From here we add additional power to the horizon equal to the power we would expect it to reflect from the sky, and from itself given some reflection coefficient, dependent on its dielectric properties.

Thus our data model reads as 

\begin{equation}
    T_D(\nu) = \frac{1}{4\pi}\int_0^{4\pi}G(\theta, \phi,\nu)\int^{t_\text{end}}_{t_\text{start}}T_\text{sky}(\theta,\phi,\nu,t)dtd\Omega + \hat{\sigma},
\end{equation}

\noindent where \(G\) is the beam pattern we generate in Section \ref{sec:beams}, \(\theta\) and \(\phi\) refer to angles over the sky, \(\nu\) is frequency, \(t\) refers to time, \(\hat{\sigma}\) is our noise parameter, and the \(T_\text{sky}\) term contains all information regarding the sky maps, CMB and horizon emission and reflection.

For a more comprehensive review of the data generation model used in this paper, see \citet{Anstey2020AExperiments} and \citet{Pattison2024ModellingOunds}.

\subsection{Bayesian Analysis}
\label{sec:bayesian}

Bayesian analysis is a statistical analysis tool derived from Bayes' theorem which aims to improve estimations of probabilities as more data becomes available such that 

\begin{equation}
    P(\Theta|D,M) = \frac{P(D|\Theta,M)P(\Theta|M)}{P(D|M)} = \frac{\mathcal{L}(\Theta)\pi(\Theta)}{\mathcal{Z}},
\end{equation}

\noindent where \(\Theta\) are the parameters of the model \(M\) we use to describe a given dataset \(D\).
In the condensed form of this equation the likelihood, \(\mathcal{L}(\Theta)\), refers to the probability of observing a dataset, given the estimated parameters and model being used to describe the data, the prior probability, \(\pi(\Theta)\), describes our antecedent knowledge of the parameters prior to sampling the data, and the evidence, \(\mathcal{Z}\) is the likelihood integrated over all parameters, weighted by the prior. 

It is through the calculation of the posterior \(P(\Theta|D,M)\) that we are able to estimate the values of the parameters from the prior space.
As this space reduces we find our posterior will iteratively become a closer match to the most probable parameter values.
The ratio of the evidences of two models will give the ratio of the probability of each model correctly fitting the data, assuming equal prior weights, based on how well the data is described by a given model.
To perform this parameter estimation and subsequent model comparison, the REACH pipeline utilises the Nested Sampling algorithm \textsc{Polychord} \citep{Handley2015PolyChord:Sampling, Handley2015Polychord:Cosmology.}.
This algorithm is given a set of prior distributions, from which it will draw a number of samples, and calculate the likelihood of said samples.
The samples with the lowest likelihoods are discarded and replaced with a new point drawn from a point of equal or higher likelihood, and the volume of the parameter space from which samples may be drawn from shrinks accordingly.
This will continue iteratively until the given termination criterion has been met.
This process produces a number of parameter estimations as well as the Bayesian evidence of the model used.
For a more in depth discussion of Nested Sampling see \citet{Skilling2006NestedComputationc} and \citet{Ashton2022NestedScientists}.

\subsection{Time-Separated Analysis Pipeline}
\label{sec:LST}

The global 21-cm signal being 5 orders of magnitude dimmer than the galactic and extragalactic foregrounds means an accurate and precise model of them is required if one wants to extract meaningful information.

This work builds upon a forward modelling method of compensating for foregrounds, originally developed in \citep{Anstey2020AExperiments} for use in the Radio Experiment for the Analysis of Cosmic Hydrogen (REACH) \citep{deLeraAcedo2022The7.528}. 
It is a physically motivated model that aims to describe the spatial variation of the foregrounds by fitting for their spectral indices.
The sky is divided into a number of regions based on the GSM 230\,MHz map \citep{deOliveira-Costa2008AGHz}, such that regions with a similar spectral index are grouped together.
One will then fit for the spectral indices of these regions, the temperature and reflection coefficient of the horizon surrounding the antenna \citep{Pattison2024ModellingOunds} and the parameters of 21 cm signal. 

With the Earth rotating on its axis through the night and around the Sun through the year the skies overhead are ever changing.
These changing skies and the resultant changing foregrounds mean any model that does not account for this time variation will be limited in its ability to extract meaningful information about the signal.
Therefore \citet{Anstey2023UseModelling} further develops this time integrated model to allow for time variation in observation, in which multiple time bins may be fit simultaneously to produce a single set of signal parameters.
It is important to note, however, that this model, while being able to account for the movement of the skies overhead assumes that all fitted parameters are constant in each time bin.
This is because if we were to let these parameters have time varying properties we would see the different parts of the data trying to converge on different posteriors simultaneously.

The likelihood\footnote{The specific likelihood used in this work differs from that in \citet{Anstey2023UseModelling} as the inclusion of a horizon complicates the foreground model.
The explicit form and formulation of this likelihood may be found in Appendix \ref{sec:likelihood}.} used for this local sidereal time pipeline may be written as:

\begin{equation}
    \begin{aligned}
        \text{log}\mathcal{L} = &\left[\sum_i\sum_j-\frac{1}{2}\text{log}(2\pi\theta_\sigma^2) \right.\\&\left. -\frac{1}{2}\left(\frac{T_D(\nu_i,t_j) - (T_F(\nu_i,t_j,\theta_F)+T_S(\nu_i,\theta_S))}{\theta_\sigma}\right)^2\right],
    \end{aligned}
    \label{eq:likelihood}
\end{equation}

\noindent where \(T_D(\nu_i,t_j)\) is the observation data, and \(T_F(\nu_i,t_j,\theta_F)\) and \(T_S(\nu_i,\theta_S)\) are the foreground and signal models, with \(\theta_\sigma\) being a Gaussian noise parameter.

This joint fit across the required number of time bins is performed using \textsc{Polychord}.
Throughout this paper this will be henceforth referred to as the `traditional pipeline'.

\begin{figure}
    \centering
    \includegraphics[width=\linewidth]{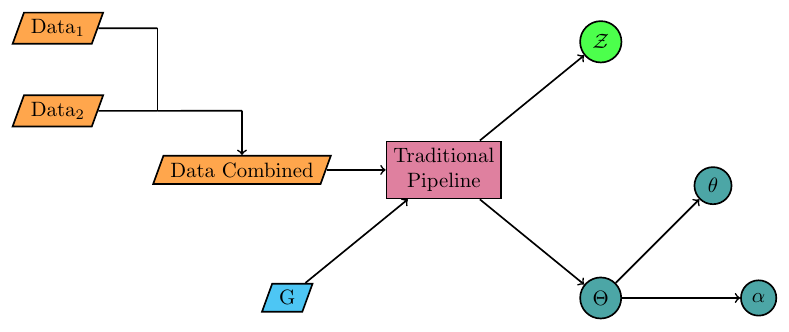}
    \caption{Flowchart showing process for two data bins going through the traditional time-separated analysis pipeline. Inputs are contained in trapeziums, a rectangle denotes a \textsc{polychord} fit, and data in circles denotes an output. The two bins of data are combined, with a series of temperatures for a given frequency across a range of time bins, these are then fitted with one beam gain pattern (G) to produce a series of nuisance and cosmological parameter estimates and an evidence for the model used.}
    \label{fig:traditional pipeline}
\end{figure}

Thus for an observation over the course of some period of time the traditional pipeline will combine a number of input data into a matrix spanning over the course of the observation, giving the antenna temperature at each frequency per unit time.
Using this data matrix, and a gain pattern of the antenna used to observe this data, we use the traditional pipeline to find a model of the sky, horizon and signal parameters and the corresponding evidences, as shown in Figure \ref{fig:traditional pipeline}.

This model must assume that each spectral index, the horizon temperature, and soil reflection coefficient remain constant over the course of an observation.
While the assumption that the spectral indices do not vary over time is a robust one, the assumption that the horizon temperature and soil reflection coefficient are unchanging is poor.
Thus we must turn to models with greater complexity to allow for this nuance.

\subsection{Normalising Flows}
\label{sec:flows}

To resolve the problem of changing environmental conditions impacting signal recovery we would want to allow these parameters to change during the fitting process.
This, however, is not feasible, allowing all time bins to have parameters independent of one another would lead to an unresolvable tension during the calculation of the posterior.
Instead, we try to perform a joint analysis using the parameters that are unchanging across time bins.
To do this we aim to perform a fit with samples that we draw from an unknown, joint distribution that we approximate with normalising flows trained on a number of simpler, known, posterior distributions.

Expressive neural networks trained to transform some generic probability distribution into a target distribution are a powerful tool for density estimation.
This work utilises the Masked Autoregressive Flows (MAFs) of \textsc{margarine} \citep{Bevins2022MarginalCosmology, Bevins2023MarginalEstimators, Bevins2023PiecewiseFlows} to perform this transformation.
An MAF is a kind of normalising flow made of a series of Masked Autoencoder for Distribution Estimation (MADE) networks \citep{Germain2015MADE:Estimation, Papamakarios2017MaskedEstimation}.

One of these networks will divide the target distribution into a number of one dimensional conditional probability distributions in which each can be approximated to be a Gaussian.

We draw samples, \(j\), from the complex distribution, \(\theta\) into our network, outputting values of the mean and standard deviation \(\mu\) and \(\sigma\), to reconstruct samples on the standard normal, \(z\), for a given dimension \(i\)

\begin{equation}
    z_i^j = \frac{\theta_i^j - \mu_i^j}{\sigma_i^j}.
\end{equation}

\noindent As the network goes through its training process the samples on the complex target distribution are transformed onto the standard normal via learnt means and standard deviation.
We aim to train the network such that the statistical distance between samples taken from the true distribution and the one produced by the network is minimised.
We do this by minimising the Kullback-Leibler divergence between the two distributions, a quantifier of this distance.
This is done by trying to maximise the log-probability predicted by the network for a set of samples on the true distribution

\begin{equation}
    \mathop{\text{argmax}}_w\sum^N_{j=0}\text{log}P_w(\theta^j),
\end{equation}

\noindent where \(P_w(\theta)\) is the probability distribution of the network for a given set of weights, \(w\).

Chaining a series of MADEs together we create our MAF, training each link in the chain simultaneously, with each previous iteration feeding into the next.
Where given a large enough architecture we can learn complex target distributions.

A more complete description of \textsc{margarine} and the MAFs it uses are found in \citet{Bevins2022MarginalCosmology, Bevins2023MarginalEstimators, Bevins2023PiecewiseFlows}
and \citet{Kobyzev2019NormalizingMethods, Papamakarios2017MaskedEstimation} respectively.

\subsection{Novel Joint Analysis Pipeline}
\label{sec:novel}

The traditional pipeline must assume that soil parameters are unchanging over time, but this is nonphysical.
While one may assume that the spectral indices of the sky regions one fits for remain constant, the noise parameter is time independent, and the parameters of the global signal are constant over the course of an observation, the parameters describing the behaviour of the soil will not be.
Over the course of a night the soil temperature may vary as much as 30K, and the soil reflection coefficient is heavily dependent on moisture level which is not constant with season.
With changing dielectric properties of the soil in the near field we observe significant changes to the expected beam pattern, which must be accounted for in our analysis of the incoming data.

\begin{figure*}
    \centering
    \includegraphics[width=\linewidth]{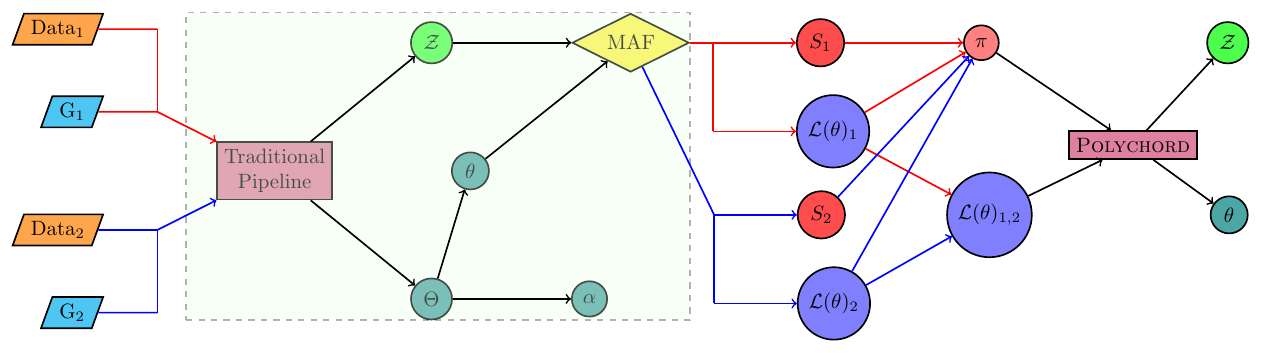}
    \caption{Flowchart showing process for two data bins going through the novel pipeline. A parallelogram represents input data, a rectangle is a \textsc{Polychord} fit, anything in a circle represents an output, and the diamond is the neural network generating normalising flows. The inputs of this pipeline are the antenna temperature data and the gain pattern (G) of the beams used. Each set of inputs are treated separately by everything in the green box, the final output being an MAF, from which we take samples (S) and a likelihood surface. The likelihood surfaces are combined as per Equation \ref{eq:likelihoodcombine}, and we use the likelihood weighted samples to form a new set of priors for the secondary \textsc{polychord} fit. This fit outputs the revised estimations of the parameters of interest and an evidence for the model.}
    \label{fig:flowflowchart}
\end{figure*}

The novel pipeline combines the traditional pipeline and \textsc{margarine} to allow for the description of these time varying parameters, while maintaining the signal parameters constraining properties of the traditional pipeline in a scenario where the environmental conditions are unchanging.
We show this process in Figure \ref{fig:flowflowchart}.
Instead of taking all our observation data and combining it for the traditional pipeline to fit with \textsc{Polychord}, we divide it into a series of binned data.

\begin{equation}
    T_D(\nu,t) = T_{D_1}(\nu,t) + T_{D_2}(\nu,t) + ... +T_{D_m}(\nu,t),
\end{equation}

\noindent where \(i\) and \(j\) refer to a 
How we determine to bin this data will depend on the tolerances of the investigation - an investigation into how overnight soil temperature variation affects signal detection will demand a finer binning than an investigation into antenna variation over series of days.

We divide our data into a series of segments, small enough that the soil properties can be approximated to be constant, but large enough that the pipeline can come near to recovering a signal over a single run.
We utilise the traditional pipeline on each of these segments to output a number of parameters \(\Theta_m\), their corresponding likelihoods, and an evidence for the overall fit.

These parameters may be further divided in an astrophysical context, we may subdivide our parameter space into nuisance and cosmological parameters of interest, giving \(\Theta = \{a,\theta\}\). 
Here we describe the signal parameters, which in the case of a Gaussian would be depth, width and centre frequency, as our cosmological parameters, and our sky region spectral indices, soil parameters and noise as nuisance parameters.
This allows us to marginalise our posterior by integrating out the nuisance parameters, \(\alpha\), such that our new, nuisance-free likelihood reads 
\begin{equation}
    \mathcal{L}(\theta) = \frac{\int\mathcal{L}(\theta,\alpha)\pi(\theta,\alpha)d\alpha}{\int\pi(\theta,\alpha)d\alpha} = \frac{\mathcal{P}(\theta)\mathcal{Z}}{\pi(\theta)},
\end{equation}
\noindent as per \citep{Bevins2022MarginalCosmology}.

We then use \textsc{margarine} to train a series of MAFs on the likelihood distribution from the \textsc{Polychord} outputs from the fits to the individual time bins.
Utilising the evidences from the original \textsc{Polychord} fit, we then sample the trained MAFs to give a likelihood weighted mean and standard deviation of each flow.
We use these values to produce a new set of priors, such that the priors for each of the parameters are defined as the likelihood weighted mean of the samples \(\pm\)5 times the standard deviation.
Finally we perform a joint fit with \textsc{Polychord}, combining all these likelihood surfaces into a new likelihood such that

\begin{equation}
    \text{log}\mathcal{L}_{1, 2, .., m}(\theta) = \text{log}\mathcal{L}_1(\theta) + \text{log}\mathcal{L}_2(\theta) + ... + \text{log}\mathcal{L}_m(\theta).
    \label{eq:likelihoodcombine}
\end{equation}

From this we will gain a description of the nuisance-free parameter space.
Finally we reduce the evidence the model outputs by the ratio of the volume of the initial set of priors used in the traditional pipeline stage, over the revised priors we input to our second \textsc{polychord} fit.

\section{Results}
\label{sec:results}
Here we discuss how the novel and traditional pipelines deal with changing environmental conditions in the near and far field.
Section \ref{sec:beamvar} describes how the two pipelines are impacted by changing dielectric constants in the nearfield soil, and Section \ref{sec:tempvar} details how they deal with changing soil temperatures in the far field.

\subsection{Beam Variability}
\label{sec:beamvar}
The form of a antenna beam is intrinsically tied to its near field environment - an antenna sitting on a more reflective soil-bed will have a greater directivity upwards, whereas one without this redirected power will look to the horizon more.
We explore here how a beam changing over a series of days due to inclement weather will affect the signal recovery using the traditional time separated analysis pipeline with respect to the novel method.

To test this we create a data set using beams where the near-field soil the antenna sits on has three different permittivities, \(\epsilon_i\), on three subsequent days.
We generate the data as described in \citet{Pattison2024ModellingOunds}, and inject a Gaussian signal with a central frequency of 85\;MHz and a width and depth of 15\;MHz and 0.155K respectively, chosen to sit well within the bounds of current theoretical models \citep{Cohen2017ChartingSignal}.
For this test we use the beam described in Section \ref{sec:beams}, that of the REACH hexagonal dipole antenna 1\,m above a bed of single layer, semi-infinite, flat soil, in the Karoo reserve.
The permittivites of this soil are such that over the series of three days the dielectric constants are equal to \(\epsilon_1 = 7.6+0.58j\), \(\epsilon_2 = 5.7+0.43j\), and \(\epsilon_3 = 3.8+0.29j\).
This kind of scenario would mimic a day of rain over a radiometer followed by two days where the soil gets gradually drier over time.
Going forwards we will refer to these as the `Wet', `Damp' and `Dry' beams respectively.
An example of the Dry gain pattern is shown in Figure \ref{fig:beampattern}.
In the signal recovery process we model the sky as having 20 different regions, the spectral indices of which will be fit for by the pipeline to recover the signal parameters.

\begin{figure}
    \centering
    \includegraphics[width =\linewidth]{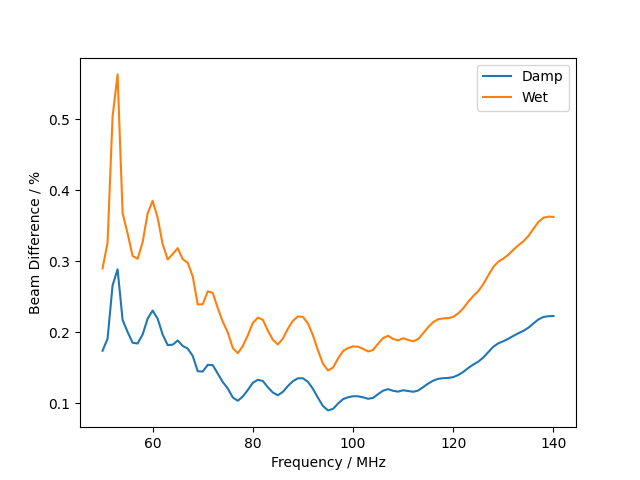}
    \caption{A comparison of generated beams using different dielectrics to describe the soil under the antenna. We show the mean absolute percentage difference between the Damp and Wet beams at each frequency compared to our Dry beam. The Wet, Damp and Dry beams have dielectric constants of \(\epsilon_1 = 7.6+0.58j\), \(\epsilon_2 = 5.7+0.43j\), and \(\epsilon_3 = 3.8+0.29j\) respectively.}
    \label{fig:beamdiff}
\end{figure}

In Figure \ref{fig:beamdiff} we show the extent to which these beam models differ. 
We plot the average absolute percentage difference between each pixel at each frequency when we compare the Damp and Wet beams to the Dry one.
This shows us that in the best case scenario, looking at the frequency channel with the least variation between beams, around 95MHz, we find an average deviation between the Damp and Dry beams, and Damp and Wet beams of approximately 1 and 1.5 parts in 1000.
This gets as bad as 3 and almost 6 parts in 1000 at about 52MHz.
This means that signal recovery with beams that change to this extent will be challenging when trying to find a signal that sits within the 70-100MHz range that theory predicts \citep{Cohen2017ChartingSignal} and even more difficult with exotic signals.

In testing the traditional pipeline we collate this data into one set, and as this pipeline will only fit our data using a single beam we feed it the Wet, Damp and Dry beams, to test how well it can recover the signal with each of them.
We test the novel pipeline by splitting this data into three separate sets according to the beam which produced them, fit them with the correct beams, and then perform a joint fit across the three outputs of the MAFs we train with the pipeline outputs.
For this test we assume that we can know precisely the form of the beam.
This is a bold assumption, as either it means we can perfectly model the soil, its moisture levels and the extent to which this impacts the gain pattern of the beam, or it means we have a tool that can fit for this beam pattern without fitting out the signal itself.
While this approximation may be an optimistic one, we believe it provides a good baseline for an exploration into the power of this novel approach.

We also create a control set, such that the beam is constant over the three days, using the Dry beam.
We fit this using the traditional model, comparing RMSEs with the novel pipeline, to show that in performing a joint fit in the novel method, we maintain the accuracy and precision of the traditional pipeline.

\begin{table*}
\caption{Table detailing signal recovery using the traditional versus novel pipeline when we have a changing antenna beam as the result of decreasing moisture levels in the soil over time. Each model described as `traditional' follows the signal recovery process described in Section \ref{sec:LST}. The moisture levels beside the model refer to an analogous dielectric constant for the beam we use in signal recovery. The novel recovery model is described in Section \ref{sec:novel}. The inserted mock signal has an 85\,MHz Central Frequency, a signal width of 15\,MHz, and a Depth of 0.155\,K. 
\(\mathcal{Z}_\text{Gauss}\) is the Bayesian evidence of trying to fit the injected signal with a Gaussian.
\(\mathcal{Z}_\text{No 21}\) is the Bayesian evidence when we try to model for our data having no 21\,cm signal.
In the traditional model this is done by fitting for only the sky regions and not a signal, in the novel model we restrict the priors in the initial fitting process such that any recovered signal has a depth an order of magnitude below noise level.
\(\delta_{\text{Log}(\mathcal{Z})}\) is the difference in evidence between these models.
RMSE refers to the root mean squared error when comparing the injected mock signal to one that we generate using the posterior averages that our Gaussian model suggests.}
\label{tab:changingbeamscomp}
\centering
\begin{tabular}{lllllllll}
\hline
&F\(_0\) (MHz) & Signal Width (MHz) & Depth (K) & Log(\(\mathcal{Z}_\text{Gauss}\)) & Log(\(\mathcal{Z}_\text{No 21}\)) & \(\delta_{\text{Log}(\mathcal{Z})}\) & RMSE (K)\\
\hline
\hline
\hline
Injected Signal & 85.0 & 15.0 & 0.155\\
\hline
Models\\
\hline
Traditional (Wet) & \(84.1\pm0.4\)& \(10.2\pm0.2\) &\(0.185\pm0.011\) & \(-1468.2\pm0.4\) & \(-1646.6\pm0.4\) & \(178.4\pm0.6\) & 0.0176\\
Traditional (Damp) &\(86.5\pm0.9\) & \(11.1\pm0.7\) & \(0.127\pm0.015\) & \(-477.3\pm0.4\) & \(-555.8\pm0.4\) & \(78.5\pm0.6\) & 0.0228\\
Traditional (Dry)  &\(84.3\pm0.9\)& \(12.1\pm0.7\)&\(0.172\pm0.022\)  &\(-903.7\pm0.4\) &\(-968.1\pm0.4\) & \(64.4\pm0.6\) & 0.0101\\
Novel & \(85.8\pm0.5\) & \(13.1\pm0.4\) & \(0.133\pm0.008\) &\(7665.7\pm0.4\) & \(7178.3\pm0.2\)& \(487.4\pm0.4\) &0.0142\\
\hline
Control & \(85.9\pm0.5\) & \(13.2\pm0.4\) & \(0.136\pm0.009\) & \(7917.9\pm0.4\) &\(7403.5\pm0.4\)&\(514.4\pm0.6\)& 0.0130\\
\hline

\end{tabular}

\end{table*}

From Table \ref{tab:changingbeamscomp} we see that the novel pipeline performs consistently better than the traditional pipeline when we have changing beams.
The novel method produces a signal model with far greater Bayesian evidence than any of the traditional models that do not account for our changing beam as the soil dries.

When we compare the novel method to the control, we can see that it performs comparably in terms of RMSE.
This shows that the novel method when performing a joint fit over three days with a correctly estimated changing beam performs as well as one might expect a traditional model to given a constant beam over the same time period.
We cannot compare the evidences of the control and novel methods as they are fits performed on different data.

\begin{figure*}
    \centering
    \begin{subfigure}[b]{0.48\textwidth}
        \centering
        \includegraphics[width=\textwidth]{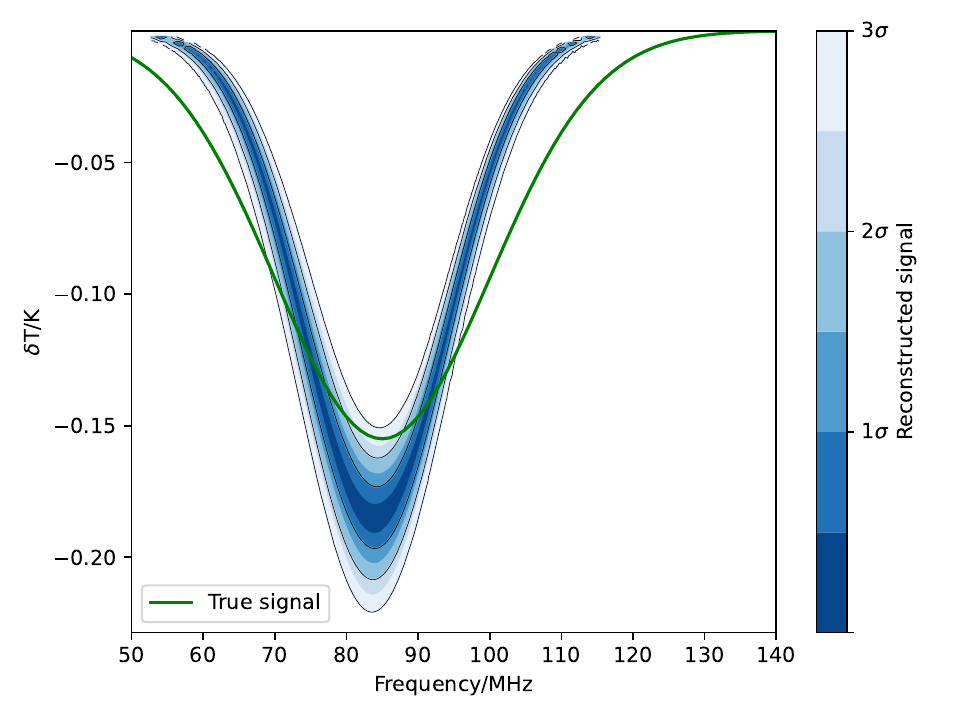}
        \caption{Signal recovery is performed using the traditional pipeline assuming the soil remains in its constant Wet state.}
        \label{fig:wet}
    \end{subfigure}
    \hfill
    \begin{subfigure}[b]{0.48\textwidth}
        \centering
        \includegraphics[width=\textwidth]{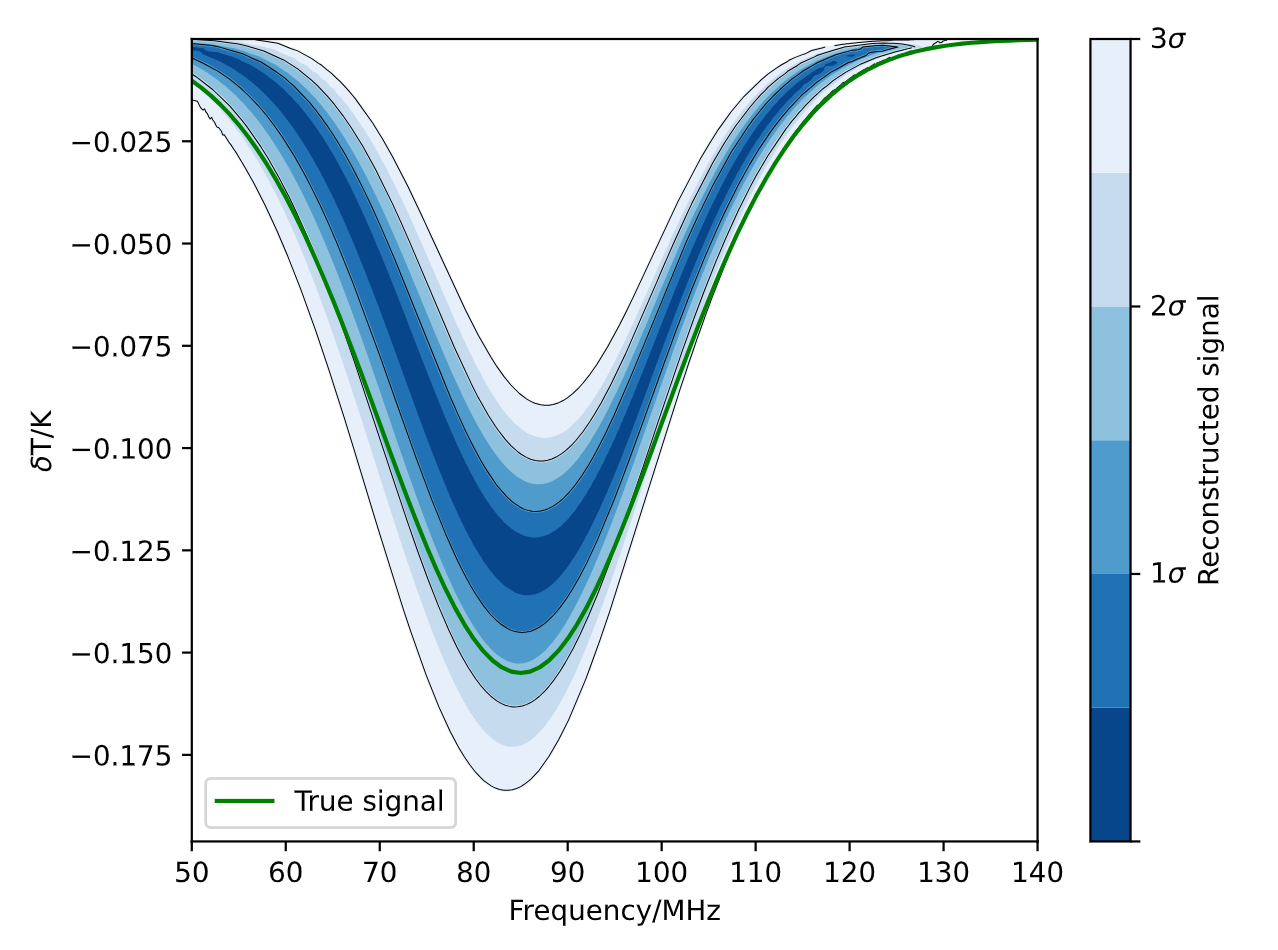}
        \caption{Signal recovery is performed using the traditional pipeline assuming the soil remains in its constant Damp state.}
        \label{fig:damp}
    \end{subfigure}
    \vskip\baselineskip
    \begin{subfigure}[b]{0.48\textwidth}
        \centering
        \includegraphics[width=\textwidth]{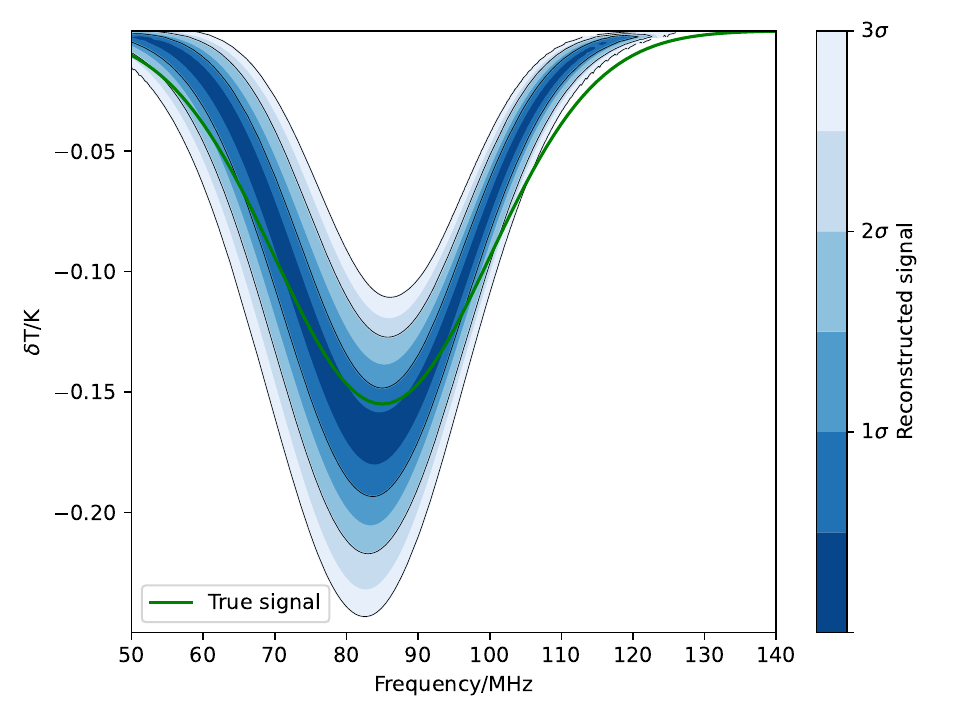}
        \caption{Signal recovery is performed using the traditional pipeline assuming the soil remains in its constant Dry state.}
        \label{fig:dry}
    \end{subfigure}
    \hfill
    \begin{subfigure}[b]{0.48\textwidth}
        \centering
        \includegraphics[width=\textwidth]{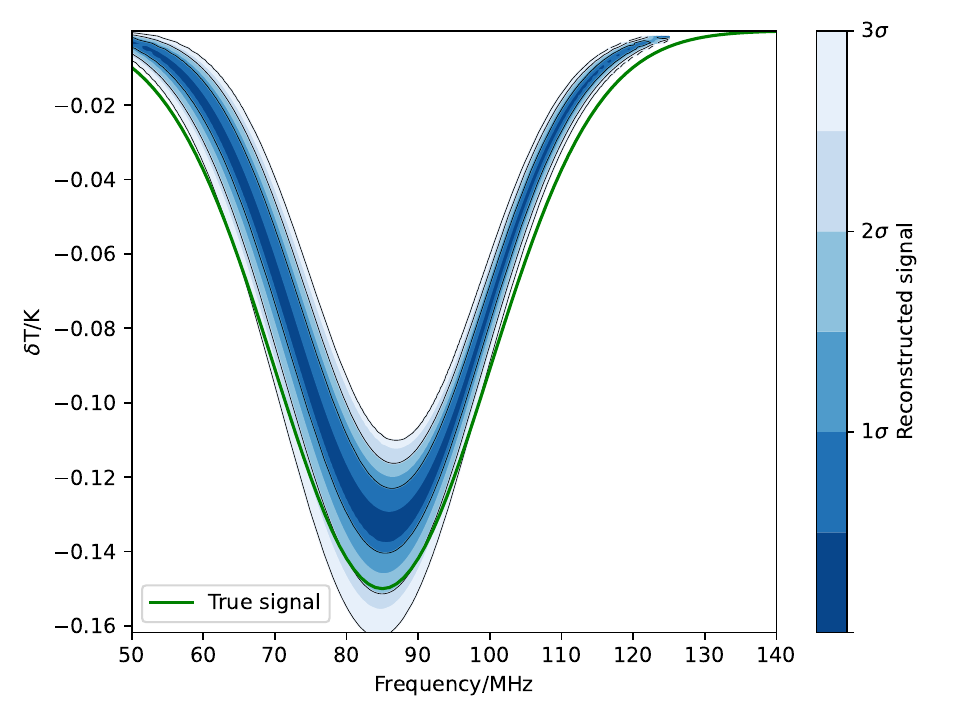}
        \caption{Signal recovery is performed using the novel pipeline where we allow for variation of soil parameters over time.}
        \label{fig:novel}
    \end{subfigure}
    \caption{Recovery of a redshifted 21\,cm signal following a 3 day mock observation in which a wet soil gets dryer over time. Injected `True' signal shown in green, with 85\,MHz Central Frequency, 15\,MHz Bandwidth, 0.155\,K Depth.}
    \label{fig:main}
\end{figure*}

Almost all methods are able to recover the signal centre frequency to within twice the quoted error bounds, but to struggle with the signal width and depth.
We see that the Wet and Dry beams in Figures \ref{fig:wet} and \ref{fig:dry} overshoot signal depth.
The Wet beam finds the depth incorrectly to 3\(\sigma\) and judges the width incorrectly to the order of 24\(\sigma\)
The traditional model using Dry beam finds the Depth to within error, and the width to \(\sim\) 4\(\sigma\) however, this has consistently the largest error bars of any of the models used.
The Damp beam and novel pipeline in Figures \ref{fig:damp} and \ref{fig:novel} both undershoot the signal depth, and though in terms of absolute magnitude the Damp beam does a worse job of finding the signal depth, the novel pipeline finds this value with a much higher precision, so is further away in terms of the error value quoted.

All models produce a positive \(\delta_\text{log(\(\mathcal{Z}\))}\), is the log of the ratio of the evidences produced when one fits a signal to the data, versus when one chooses to just fit for the foregrounds.
This indicates that for all of them, even though recovery is not perfect, we would favour some kind of a detection over a non detection if working with real data.
We note here that in the case of the novel pipeline, we describe a fit for just the foregrounds involves fitting for a signal with priors below the level of the noise we inject into our data. 
We do this as in order to learn the distribution of our signal with \textsc{margarine} we need to provide it with data to learn from. 
If we solely fit for the foregrounds in the initial stages of the pipeline, there will be no information to work from, so we have no way to compare a set of data with a signal to one without.
Restricting the priors to be such that the signal must be below the noise level means that the signal is not recovered, and provides a good surrogate for a foreground only fit.

If one looks solely at the RMSE of the recovered signal compared to the injected one, one might assume that the traditional pipeline recovery that uses the Dry beam in signal recover performs the best.
However in practice it has an evidence lower than the traditional method using the Damp beam, and an evidence substantially lower than the novel method.
As a result, if this test were to be performed on real data the odds of our novel pipeline giving us the correct signal model would be far higher.

This test is performed not only to demonstrate the capabilities of the novel pipeline, but also to check how the traditional one would deal with recovering a non-exotic 21\;cm signal with a changing beam.
As the signal becomes more exotic, and drifts further into the regions below 70MHz or above 120MHz we would see the novel pipeline becoming even more powerful as the difference between beams increases further.

\subsection{Soil Temperature Variation}
\label{sec:tempvar}

Deserts are some of the most extreme environments in the world, where soil temperatures can range between \(80^\circ\)C \footnote{There have been reports that natural soil temperatures have reached up to \(93.9^\circ\)C \citep{Kubecka2001ATemperature} in Death Valley but these claims are not confirmed.} \citep{Handley1970MicrometeorologyArthropods} and near freezing.
Deserts, such as the Karoo, Inyarrimanha Ilgari Bundara, and Death Valley have been home to the REACH, EDGES and MIST \citep{Monsalve2024MapperOverview} radiometers, and so we use these as our testing ground for the novel pipeline.

In this section we investigate how the novel and traditional pipelines are able to recover the 21\,cm signal in the face of changing soil temperatures, over the course of one evening, and over a year, using the Karoo radio reserve as our test case.

\begin{figure}
    \centering
    \includegraphics[width = \linewidth]{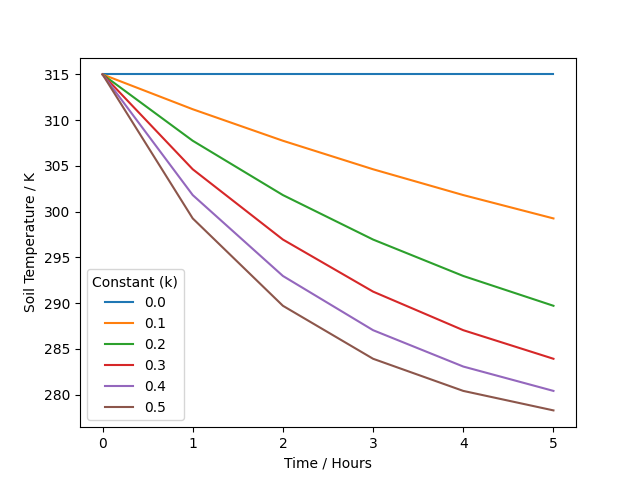}
    \caption{Soil temperature over the course of an evening where each coloured line represents a soil temperature modelled following Equation \ref{eq:cooling} where the corresponding cooling coefficient for each soil is shown in the legend.}
    \label{fig:soiltemp}
\end{figure}

\begin{figure*}
    \hspace*{-1.5cm}
    \centering
    \includegraphics[width = \textwidth]{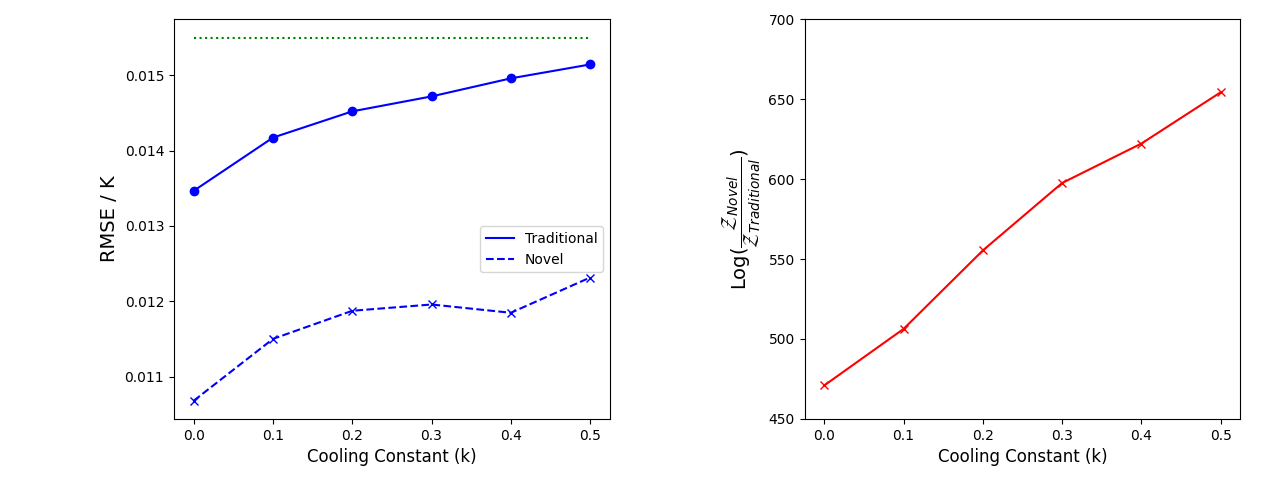}
    \caption{Plots comparing how the Bayesian evidence and RMSE of signal recovery with the traditional and novel pipeline change over a number of observations with changing soil temperatures. The plot on the left shows the RMSE of the recovered signals for each cooling constant, where the RMSE of the traditional pipeline is shown with a solid line, and the RMSE of the signal recovered with the novel pipeline is shown with a dashed line. The green dotted line indicates an RMSE of one tenth the depth of the  injected signal. The Right hand plot shows the log of the ratio of evidences of signal recovery between the novel and traditional pipeline for a dataset with a given cooling constant.}
    \label{fig:overnightcomp}
\end{figure*}

\subsubsection{Overnight Temperature Variation}

Here we present the results of letting the soil temperature on the horizon vary over the course of 6 hours.
We test models with soil changes ranging from an extreme example, to something more realistic, to no variation at all over the course of an evening in order to compare the efficacy of the traditional versus novel pipelines dealing with shifting fitted parameters.

We model the temperature drop overnight as an exponential decay following Newton's law of cooling \citep{Newton1701VII.Caloris}, where we assume a constant ambient temperature and a convective heat loss, with

\begin{equation}
    T_H = T_a + 40 e^{(-kt)},
\label{eq:cooling}
\end{equation}

\noindent where \(T_H\) is the temperature of the soil on the horizon, \(T_a\) is the air temperature after sundown, \(k\) a cooling rate constant, and \(t\) is the time in hours from midnight. 
For this test we approximate the air temperature to be 273K after sundown, and vary \(k\) between 0 and 0.5 to give examples of slow and rapid temperature reductions over an evening, with k = 0 being a control in which there is no temperature variation.
While a more realistic soil temperature model could be found by specific examination of the overnight temperature variation in the Karoo these approximations are used to stress-test the novel pipeline, as opposed to being an exact physical model.
This is shown in Figure \ref{fig:soiltemp}.
In the case of the traditional pipeline we observe for the full 5 hours, with a new time bin every 5 minutes; the novel pipeline subdivides this into hour long segments with 5 minute time bins, these outputs are then jointly fit as described in Section \ref{sec:novel}.

From Table \ref{tab:overnightcomp} and Figure \ref{fig:overnightcomp} we can see how the traditional and novel methods compare.
Both methods find the signal to a high degree of accuracy, with an RMSE corresponding to less than a tenth of the signal depth.
However, we can see that both methods show a general trend of increasing RMSE with an increased rate of cooling - the novel method is able to maintain a consistent RMSE for longer, but sees a spike between \(k = 0.4\) and \(k=0.5\).
Despite both being subject to a higher RMSE at high \(k\) values, the novel method recovers a signal consistently closer to the one we inject, with an average RMSE 0.0028\,K lower than the traditional one, corresponding to a 20\% decrease.

The Bayesian evidence of the signal recovery in the traditional pipeline also notably decreases as \(k\) increases with respect to the novel pipeline, which remains mostly consistent with changing \(k\).
We see this is Figure \ref{fig:overnightcomp} where the log of the ratio of the evidences of the novel pipeline versus the traditional one increases dramatically as we increase the cooling constant.
The ratio of these evidences indicate the odds that one model is correct when compared to the other, so this means that as we increase the cooling constant the odds that the novel pipeline produces a more correct model than the traditional pipeline rises quickly.

\begin{table*}
\caption{Table comparing signal recovery over a series of mock observations where the temperature of the far field soil varies according to Equation \ref{eq:cooling} where \(k\) describes the cooling constant used in this equation.
The novel recovery model is described in Section \ref{sec:novel}.
The traditional model follows the signal recovery process described in Section \ref{sec:LST}. 
 The inserted mock signal has an 85\,MHz Central Frequency, a signal width of 15\,MHz, and a Depth of 0.155\,K. 
\(\mathcal{Z}_\text{Gauss}\) is the Bayesian evidence of trying to fit the injected signal with a Gaussian.
\(\mathcal{Z}_\text{No 21}\) is the Bayesian evidence when we try to model for our data having no 21\,cm signal.
In the traditional model this is done by fitting for only the sky regions and not a signal, in the novel model we restrict the priors in the initial fitting process such that any recovered signal has a depth an order of magnitude below noise level.
\(\delta_{\text{Log}(\mathcal{Z})}\) is the difference in evidence between these models.
RMSE refers to the root mean squared error when comparing the injected mock signal to one that we generate using the posterior averages that our Gaussian model suggests. }
\label{tab:overnightcomp}
\centering
\begin{tabular}{lllllllll}
\hline
&F\(_0\) (MHz) & Signal Width (MHz) & Depth (K) & Log(\(\mathcal{Z}_\text{Gauss}\)) & Log(\(\mathcal{Z}_\text{No 21}\)) & \(\delta_{\text{Log}(\mathcal{Z})}\) & RMSE (K)\\
\hline
\hline
\hline
Injected Signal & 85.0 & 15.0 & 0.155\\
\hline

\(k=0.0\)\\
novel  &  \(85.4\pm0.3\)& \(13.4\pm0.3\) & \(0.140\pm0.006\) & \(15355.5\pm0.3\)& \(14514.6\pm0.3\) & \(840.9\pm0.4\) & 0.0107\\
Traditional & \(85.7\pm0.4\) & \(13.3\pm0.3\) & \(0.134\pm0.006\) & \(14884.6\pm0.9\)& \(13998.8\pm0.5\)& \(885.8\pm1.0\) & 0.0135\\
\hline
\(k=0.1\)\\
Novel & \(85.5\pm0.4\) & \(13.3\pm0.3\) & \(0.138\pm0.006\) & \(15349.6\pm0.3\) & \(14511.6\pm0.3\) &  \(838.0\pm0.4\)& 0.0115\\
Traditional & \(85.8\pm0.4\) & \(13.2\pm0.3\) & \(0.133\pm0.007\)& \(14843.3\pm0.4\) &\(13951.7\pm0.5\)& \(891.6\pm0.6\) & 0.0142\\
\hline
\(k=0.2\)\\
Novel & \(85.5 \pm0.4\) & \(13.3\pm0.3\) & \(0.138 \pm0.006\) & \(15345.5\pm0.3\) &\(14354.1\pm0.3\)&\(991.4\pm0.4\)& 0.0115 \\
Traditional & \(85.9\pm0.4\) & \(13.1\pm0.3\) & \(0.132\pm0.006\)& \(14790.1\pm0.5\) & \(13897.9\pm0.5\) & \(892.2\pm0.7\)& 0.0145\\
\hline
\(k=0.3\)\\
Novel & \(85.5\pm0.3\) & \(13.3\pm0.3\)  & \(0.138\pm0.006\) & \(15342.3\pm0.3\) &\(14493.1\pm0.3\)& \(849.2\pm0.4\)& 0.0119\\
Traditional & \(85.9\pm0.4\) & \(13.1\pm0.3\) & \(0.132\pm0.006\)& \(14744.7\pm0.5\)&\(13860.5\pm0.5\)& \(884.2\pm0.7\) & 0.0147\\
\hline
\(k=0.4\)\\
Novel & \(85.6 \pm0.3\) & \(13.3 \pm0.3\) & \(0.138\pm0.006\) & \(15337.4\pm0.3\) & \(14491.5\pm0.3\) &\(845.9\pm0.4\)& 0.0118\\
Traditional & \(85.9\pm0.4\) & \(13.0\pm0.3\) & \(0.132\pm0.006\) & \(14715.1\pm0.4\) &\(13834.1\pm0.5\) & \(881.0\pm0.6\) & 0.0150\\
\hline
\(k=0.5\)\\
Novel & \(85.6 \pm0.3\)& \(13.2\pm0.3\) & \(0.137\pm0.006\) & \(15343.1\pm0.3\) & \(14487.4\pm0.3\)& \(855.7\pm0.4\) & 0.0123\\
  
Traditional & \(86.0\pm0.4\) & \(13.0\pm0.3\) & \(0.131\pm0.006\) & \(14688.7\pm0.4\) & \(13810.1\pm0.5\)& \(878.6\pm0.6\)&0.0151\\

\hline

\end{tabular}

\end{table*}

\subsubsection{Seasonal Temperature Variation}
\label{sec:seasonal}

Over the course of a year the average temperature in the Karoo varies by about 10K, at a minimum in July and reaching peaks in January.
We investigate how the novel and traditional pipelines are able to deal with this variation by feeding them mock data from observations over the course of a year.
We set the temperature over the course of a night to be constant: in January to 290K, in April and October to be 285K and in July to be 280K.

To do this we make four mock data sets, each four hours long with an observation start time as the 1st of each month at midnight.
Each of these observations has four one hour time bins.
As in the case of the overnight variation, we let the traditional pipeline fit all this data at once, but we let the novel pipeline take each of these days individually to be fit jointly after training the MAFs.

\begin{table*}
\caption{Table comparing signal recovery over a series of mock observations where the temperature of the far field soil varies over the course of a year as described in Section \ref{sec:seasonal}.
The novel recovery model is described in Section \ref{sec:novel}.
The traditional model follows the signal recovery process described in Section \ref{sec:LST}. 
 The inserted mock signal has an 85\,MHz Central Frequency, a signal width of 15\,MHz, and a Depth of 0.155\,K. 
\(\mathcal{Z}_\text{Gauss}\) is the Bayesian evidence of trying to fit the injected signal with a Gaussian.
\(\mathcal{Z}_\text{No 21}\) is the Bayesian evidence when we try to model for our data having no 21\,cm signal.
In the traditional model this is done by fitting for only the sky regions and not a signal, in the novel model we restrict the priors in the initial fitting process such that any recovered signal has a depth an order of magnitude below noise level.
\(\delta_{\text{Log}(\mathcal{Z})}\) is the difference in evidence between these models.
RMSE refers to the root mean squared error when comparing the injected mock signal to one that we generate using the posterior averages that our Gaussian model suggests. }

\centering
\begin{tabular}{lllllllll}
\hline
&F\(_0\) (MHz) & Signal Width (MHz) & Depth (K) & Log(\(\mathcal{Z}_\text{Gauss}\)) & Log(\(\mathcal{Z}_\text{No 21}\)) & \(\delta_{\text{Log}(\mathcal{Z})}\) & RMSE (K)\\
\hline
\hline
\hline
Injected Signal & 85.0 & 15.0 & 0.155\\
\hline

Novel & \(86.6\pm0.3\) & \(12.7\pm0.4\) & \(0.125\pm0.004\) & \(3535.0\pm0.3\) &\(2965.5\pm0.4\)&\(569.6\pm0.5\)&0.0187\\
Traditional & \(88.0\pm0.3\) & \(10.3\pm0.2\) & \(0.010\pm0.002\) & \(3447.3\pm1.1\) & \(2867.9\pm0.5\) & \(579.4\pm1.2\) & 0.0337\\

\hline

\end{tabular}
\label{tab:seasonal}
\end{table*}

We see from Table \ref{tab:seasonal} that the novel pipeline once again is able to recover the signal to a higher degree of accuracy than the traditional one.
The novel pipeline recovers the injected signal with an RMSE 45\% lower than that of the traditional pipeline, with a higher Bayesian evidence.

\section{Conclusions}
\label{sec:conclusions}
We demonstrate in this paper that a traditional physically motivated Bayesian analysis pipeline like the one designed for REACH will struggle to account for variation in soil moisture levels and the resultant changes to the beam.
Thus, the novel pipeline is implemented.
This takes the data we get from our antenna and divides it into a number of time bins where we can assume that the soil properties are unchanging.
It fits them with the traditional pipeline and trains a series of MAFs from which we take samples to then perform a joint fit across all samples using the \textsc{polychord} nested sampling algorithm to better recover the redshifted 21\,cm signal parameters.
We allowing one to fit for signal parameters using changing beams over time and then using a neural network to collate these data and perform a joint fit across them.

We find that the novel pipeline outperforms the traditional one in the presence of a changing antenna beam.
The traditional pipeline can only use one beam model in the fitting process, and with the novel pipeline not having this limitation we see a large improvement.
We see that the novel pipeline recovers the injected signal with a log evidence between 7700 and 8800 greater than any of the traditional models, with an average RMSE decrease of 15\%.
The novel pipeline performs as well with changing beams as we would expect the traditional one to perform with a consistent one, with a comparable RMSE for a given injected signal.

However, the limitation in practice that this pipeline now faces is in working out how soil moisture levels directly impact the dielectrics of a given soil and the subsequent beam.
We show in this paper that fitting for our data with a beam that is partially correct over the duration of an observation with a changing beam can increase the RMSE by 75\%.
Using a beam that is consistently incorrect, whether due to a misunderstanding of dielectric properties, or the topography of the terrain would dramatically decrease likelihood of accurate signal recovery.
This can be accounted for by either direct measurement and weather monitoring, or allowing for a parameterized beam fitting.
Direct measurement of  these soil properties would need to be done on a site by site basis; one needs to take into account moisture draining and retention based on soil types and how soil layers stack on top of one another under the antenna to give a comprehensive understanding of beam evolution over time.
A more powerful approach may be to incorporate fitting for the beams gain pattern within the data analysis pipeline.
As discussed in \citet{Cumner2023TheExperiment} one can use singular value decomposition to construct a beam pattern using numerical basis functions and fit for these basis functions within the general signal recovery process.
While this would add between 30 and 50 dimensions to our likelihood, it could prove to be an enormous improvement in our ability to work through beam uncertainties.

We also discuss how this technique also allows for the pipeline to deal with changing far-field soil temperatures over the course of one night of observation, or changes in average temperature that appear over the course of several months.
We show that while the effect of changing farfield soil parameters over these time periods does not have an impact on the ability of the traditional pipeline to recover a signal, we show that it does bias the recovery with respect to the injected signal.
The novel pipeline works to recover a signal much closer to that which was injected, showing a 20\% RMSE decrease on average over all tests where we let soil temperature change overnight, and an almost 50\% RMSE decrease when we tested seasonal temperature variation.

\section*{Acknowledgements}

We like to thank Will Handley for his integral contributions to the REACH pipeline, Christophe Craeye for his work on the numerical simulations of the REACH dipole antenna, and Laura Bartolomei-Hill for her contributions to Crayola LLC.

JHNP, DJA, and EdLA were supported by the Science and Technology Facilities Council.
HTJB acknowledges support by the Kavli Institute for Cosmology Cambridge and the Kavli Foundation.
We would also like to thank the Kavli Foundation for their support of REACH.

%%%%%%%%%%%%%%%%%%%%%%%%%%%%%%%%%%%%%%%%%%%%%%%%%%
\section*{Data Availability}

The data that support the findings of this study are available from the first author upon reasonable request.

%%%%%%%%%%%%%%%%%%%% REFERENCES %%%%%%%%%%%%%%%%%%

% The best way to enter references is to use BibTeX:

\bibliographystyle{mnras}
\bibliography{references} % if your bibtex file is called example.bib

% Alternatively you could enter them by hand, like this:
% This method is tedious and prone to error if you have lots of references
%\begin{thebibliography}{99}
%\bibitem[\protect\citeauthoryear{Author}{2012}]{Author2012}
%Author A.~N., 2013, Journal of Improbable Astronomy, 1, 1
%\bibitem[\protect\citeauthoryear{Others}{2013}]{Others2013}
%Others S., 2012, Journal of Interesting Stuff, 17, 198
%\end{thebibliography}

%%%%%%%%%%%%%%%%%%%%%%%%%%%%%%%%%%%%%%%%%%%%%%%%%%

%%%%%%%%%%%%%%%%% APPENDICES %%%%%%%%%%%%%%%%%%%%%

\appendix

\section{Likelihood Formulation}

\label{sec:likelihood}

The addition of a horizon to a physically motivated foreground model such that one can fit for soil temperature and reflection coefficient will complicate the likelihood used to recover the signal.
In this appendix we describe the formulation and form of the likelihood we use for signal recovery in a time separated physically motivated Bayesian data analysis pipeline.

As in Equation \ref{eq:likelihood}, taken from \citet{Anstey2023UseModelling}, the likelihood takes the generalised form:

\begin{equation}
    \begin{aligned}
        \text{log}\mathcal{L} = &\left[\sum_i\sum_j-\frac{1}{2}\text{log}(2\pi\theta_\sigma^2) \right.\\&\left. -\frac{1}{2}\left(\frac{T_D(\nu,t) - (T_F(\nu,t,\theta_F)+T_S(\nu,\theta_S))}{\theta_\sigma}\right)^2\right],
    \end{aligned}
    \label{eq:genlike}
\end{equation}

\noindent in which the indices \(i\) and \(j\) refer to the time and frequency bins.
As in Section \ref{sec:LST}, \(T_D(\nu,t)\) refers to the observation data, with \(T_F(\nu,t,\theta_F)\) and \(T_S(\nu,\theta_S)\) are the foreground and signal models respectively, and \(\theta_\sigma\) being a Gaussian noise parameter.

The first term in Equation \ref{eq:genlike} is  entirely independent of \(i\) and \(j\) and thus may be calculated outside the likelihood to give: \(-\frac{1}{2}N_tN_\nu\text{log}(2\pi\theta^2_\sigma)\) where \(N_t\) and \(N_\nu\) are the known constants, being number of time and frequency bins respectively.

Putting this to the side for now we focus on the second part of this equation, beginning by restating this section of our likelihood as \(\mathcal{L}_{sq}\), and expanding it, allowing us to define

\begin{equation}
    \begin{aligned}
        L_{sq} &= \frac{1}{2\theta_\sigma^2}\left(\left[\sum_i\sum_jT_{D_{i,j}}^2\right] + \left[\sum_i\sum_jT_{F_{i,j}}(\theta_F)^2\right] +\right.\\&\;\left[\sum_i\sum_jT_{S_i}(\theta_S)^2\right] -2 \left[\sum_i\sum_jT_{D_{i,j}}T_{F_{i,j}}(\theta_F)\right] \\&\;\left.-2\left[\sum_i\sum_jT_{D_{i,j}}T_{S_i}(\theta_S)\right] +2\left[\sum_i\sum_jT_{F_{i,j}}(\theta_F)T_{S_i}(\theta_S)\right]\right).
    \end{aligned}
\end{equation}

For ease of parsing we then assign the variables \(Z, Y, X, W, V, \text{and } U\) to this equation, such that:

\begin{equation}
    L_{sq} = \frac{1}{2\theta_\sigma^2}(\textcolor{smar}{Z} + \textcolor{smar}{Y} + \textcolor{smar}{X} - 2\textcolor{smar}{W} - 2\textcolor{smar}{V} + 2\textcolor{smar}{U}),
    \label{eq:lsq}
\end{equation}

\noindent where throughout this formulation, all final, fully expanded versions of these terms are written in \textcolor{smar}{blurple}.

We may then expand and evaluate these variables to allow us to maximise the efficiency of the likelihood, calculating what terms we can outside of the likelihood.

We first discuss the terms that do not include \(T_{F_{i,j}}\), where our terms will be entirely consistent with \citep{Anstey2023UseModelling}:
\begin{itemize}
    \item  \(\textcolor{smar}{Z} = \sum_i\sum_jT_{D_{i,j}}^2\), is entirely parameter independent, and is thus a constant, able to be calculated outside of the likelihood.
    \item \(T_{S_i}(\theta_S)\) is time independent, and thus \(X = \sum_i\sum_jT_{S_i}(\theta_S)^2\) may be defined as \(\textcolor{smar}{X} = N_t\sum_iT_{S_i}(\theta_S)^2\).
    \item If one defines \(\sum_jT_{D_{i,j}} = T_{D_i}\) then this value may be precalculated, letting \(\textcolor{smar}{V} = \sum_i\sum_jT_{D_{i,j}}T_{S_i}(\theta_S) = \sum_iT_{D_i}T_{S_i}(\theta_S)\).
\end{itemize}

The remaining terms are all dependent on the foreground model, \(T_{F_{i,j}}\), which, when we account for the horizon as in \citet{Pattison2024ModellingOunds} becomes:

\begin{equation}
    \begin{aligned}
        T_{F_{i,j}} = &\sum_k K_{i,j,k}F_i(\theta_{F_k}) 
        + \sum_k R_{i,j,k}F_i(\theta_{F_k})|\Gamma|\alpha \\&+ J_iT_H\left(1 + |\Gamma|\alpha\right).
    \end{aligned}
    \label{eq:Fstuff}
\end{equation}

\noindent Here \(k\) indexes the sky regions in the foreground model.
In \citet{Anstey2023UseModelling} \(K_{i,j,k}\) and \(F_i(\theta_{F_k})\) are the parameter independent, time dependent term and the parameter dependent time independent term.
Following \citet{Pattison2024ModellingOunds}, with the addition of the horizon around a radiometer \(K_{i,j,k}\) specifically becomes a chromaticity term, dealing with power coming from the sky directly; \(R_{i,j,k}\) describes power reflected from the horizon.
\(J_i\) is the analogous horizon `chromaticity' term, dealing with thermal power originating from the horizon itself, \(|\Gamma|\) is the magnitude of the reflection coefficient of the soil, \(\alpha\) is defined as \(\frac{180-\xi}{180}\) where \(\xi\) is the angle the horizon makes relative to the plane of the basin, and \(T_H\) is the soil temperature of the horizon.

We then consider the \(W \text{and } U\) terms:
\begin{equation}
    \begin{aligned}
        W &= \sum_i\sum_jT_{D_{i,j}}T_{F_{i,j}}(\theta_F)\\& = \sum_i\sum_j\sum_kT_{D_{i,j}}K_{i,j,k}F_i(\theta_{F_k}) \\&\;+ \sum_i\sum_j\sum_k T_{D_{i,j}}T_{R_{i,j,k}}F_i(\theta_{F_k})|\Gamma|\alpha \\&\;+ \sum_i\sum_j\sum_kT_{D_{i,j}}J_iT_H\left(1 + |\Gamma|\alpha\right), 
    \end{aligned}
\end{equation}

\noindent in which we define \(\sum_jT_{D_{i,j}}K_{i,j,k} = T_{D\cdot K_{i,k}}\), and similarly \(\sum_jT_{D_{i,j}}R_{i,j,k} = T_{D\cdot R_{i,k}}\); also defining \(\sum_jT_{D_{i,j}} = \sum_iT_{D_i}\) means these three terms can be calculated outside the likelihood to speed up computation time to give:

\begin{equation}
    \begin{aligned}
        \textcolor{smar}{W} = & \sum_i\sum_kT_{D\cdot K_{i,k}}F_i(\theta_{F_k}) + \sum_i\sum_kT_{D\cdot R_{i,k}}F_i(\theta_{F_k})|\Gamma|\alpha  \\& + \sum_iT_{D_i}J_iT_H\left(1 + |\Gamma|\alpha\right).
    \end{aligned}
\end{equation}    

\noindent When considering the \(U\) term:
\begin{equation}
    \begin{aligned}
        U &= \sum_i\sum_jT_{S_i}(\theta_s)T_{F_{i,j}}(\theta_F) \\&=\sum_i\sum_j\sum_kT_{S_i}(\theta_s)K_{i,j,k}F_i(\theta_{F_k}) \\&\;+ \sum_i\sum_j\sum_k T_{S_i}(\theta_s)T_{R_{i,j,k}}F_i(\theta_{F_k})|\Gamma|\alpha \\&\;+ \sum_i\sum_j\sum_kT_{S_i}(\theta_s)J_iT_H\left(1 + |\Gamma|\alpha\right),
    \end{aligned}
\end{equation}

\noindent in which we can see with rearrangement that 
\begin{equation}
    \begin{aligned}
        \sum_i\sum_j\sum_k&T_{S_i}(\theta_s)K_{i,j,k}F_i(\theta_{F_k}) =\\& \sum_i\sum_k\left(\sum_jK_{i,j,k}\right)T_{S_i}(\theta_s)F_i(\theta_{F_k}),
    \end{aligned}
\end{equation}

\noindent so if we define \(\sum_jK_{i,j,k} = K_{i,k}\), and similarly \(\sum_jR_{i,j,k} = R_{i,k}\), both of which can be precalculated we yield the equation for U:

\begin{equation}
    \begin{aligned}
        \textcolor{smar}{U} &= \sum_i\left(\sum_kK_{i,k}F_i(\theta_{F_k})\right)T_{S_i}(\theta_k) \\&\;+ \sum_i\left(\sum_kR_{i,k}F_i(\theta_{F_k})\right)T_{S_i}(\theta_k)|\Gamma|\alpha \\&\;+ N_t\Sigma_iJ_iT_H(1 + |\Gamma|\alpha)T_{S_i}F_i(\theta_k)
    \end{aligned}
\end{equation}

Finally, we consider the \(Y\) term.
To do this we will once again redefine terms from Equation \ref{eq:Fstuff}, letting \(G = \sum_k K_{i,j,k}F_i(\theta_{F_k})\), \(H = \sum_k R_{i,j,k}F_i(\theta_{F_k})|\Gamma|\left(\alpha\right)\), and \(I = J_iT_H|\Gamma|\left(1 + \alpha\right)\), such that:

\begin{equation}
    \textcolor{smar}{Y} = \sum_i\sum_jT_{F_{i,j}}^2 = \textcolor{red}{G^2} + \textcolor{red}{H^2} + \textcolor{red}{I^2} + \textcolor{red}{2GH} + \textcolor{red}{2GI} + \textcolor{red}{2HI},
\end{equation}

\noindent in which, analogous to \ref{eq:lsq}, all fully expanded versions of these terms are written in \textcolor{red}{razzmatazz}.
We begin our understanding of this equation by expanding \(G^2\):

\begin{equation}
    \begin{aligned}
        G^2 &= \sum_i\sum_j\left[\sum_k K_{i,j,k}F_i(\theta_{F_k})\right]^2 \\&= \sum_i\sum_k\left[\left[\sum_jK_{i,j,k}^2\right]F_{i,k}(\theta_{F_k})^2 \right.\\&\;
        \left.+ \sum_i\sum_{k_1 \neq k_2}\left[K_{i,j,k_1}K_{i,j,k_1}\right]F_{i,K_1}(\theta_{F_{K_1}})F_{i,K_2}(\theta_{F_{K_2}})\right],
    \end{aligned}
\end{equation}

We will then define \(\sum_jK_{i,j,k}^2 = K_{\boxdot i,k}\) and \(\sum_jK_{i,j,k_1}K_{i,j,k_2} = K_{\dagger i,k_1,k_2}\), terms which may be calculated outside the likelihood.
Thus we may define \(G^2\) as:

\begin{equation}
    \begin{aligned}
        \textcolor{red}{G^2} &= \sum_i\sum_k\left[\left[K_{\boxdot i,k}\right]^2F_{i,k}(\theta_{F_k})^2 \right.\\&\; + \sum_{k_1}\sum_{k_2} \sum_i K_{\dagger i,k_1,k_2}F_{i,K_1}(\theta_{F_{K_1}})F_{i,K_2}(\theta_{F_{K_2}}) \\&\;- \text{Tr}\sum_iK_{\dagger i,k_1,k_2})F_{i,K_1}(\theta_{F_{K_1}}F_{i,K_2}(\theta_{F_{K_2}}).
    \end{aligned}
\end{equation}

As \(R_{i,j,k}\) is analogous to \(K_{i,j,k}\) we can expand \(H^2\) in the same way such that:

\begin{equation}
    \begin{aligned}
        \textcolor{red}{H^2} &= \left(|\Gamma|\alpha\right)^2\sum_i\sum_k\left[\left[R_{\boxdot i,k}\right]^2F_{i,k}(\theta_{F_k})^2 \right.\\&\; + \sum_{k_1}\sum_{k_2} \sum_i R_{\dagger i,k_1,k_2}F_{i,K_1}(\theta_{F_{K_1}})F_{i,K_2}(\theta_{F_{K_2}}) \\&\;- \text{Tr}\sum_iR_{\dagger i,k_1,k_2}F_{i,K_1}(\theta_{F_{K_1}})F_{i,K_2}(\theta_{F_{K_2}}).
    \end{aligned}
\end{equation}

We continue this process to expand \(GH\), where we allow \(\sum_jK_{i,j,k}R_{i,j,k} = RK_{\boxdot i,k}\) and \(\sum_jK_{i,j,k_1}R_{i,j,k_2} = RK_{\dagger i,k_1,k_2}\), once again, these terms being able to be calculated outside the likelihood, to give: 

\begin{equation}
    \begin{aligned}
        \textcolor{red}{GH} &= \left(|\Gamma|\alpha\right)^2\sum_i\sum_k\left[\left[RK_{\boxdot i,k}\right]^2F_{i,k}(\theta_{F_k})^2 \right.\\&\; + \sum_{k_1}\sum_{k_2} \sum_i RK_{\dagger i,k_1,k_2}F_{i,K_1}(\theta_{F_{K_1}})F_{i,K_2}(\theta_{F_{K_2}}) \\&\;- \text{Tr}\sum_iRK_{\dagger i,k_1,k_2}F_{i,K_1}(\theta_{F_{K_1}})F_{i,K_2}(\theta_{F_{K_2}}).
    \end{aligned}
\end{equation}

\noindent The final three terms then expand to give:

\begin{equation}
    \textcolor{red}{I^2} = \left[T_H(1+|\Gamma|\alpha)\right]^2N_t\sum_iJ_i^2,
\end{equation}

\begin{equation}
    \textcolor{red}{GI} = T_H(1+|\Gamma|\alpha)\sum_i\left(\sum_kK_{i,k}J_iF_i(\theta_{F_k})\right),
\end{equation}

\begin{equation}
    \textcolor{red}{HI} = T_H(|\Gamma|\alpha+\left(|\Gamma|\alpha\right)^2)\sum_i\left(\sum_kR_{i,k}J_iF_i(\theta_{F_k})\right),
\end{equation}

\noindent leaving our final likelihood equation to read as:

\begin{equation}
    \begin{aligned}
        \text{log}\mathcal{L} = -\frac{1}{2}N_tN_\nu(2\pi\theta_\sigma^2) - \frac{1}{2\theta_\sigma^2}(\textcolor{smar}{Z} + \textcolor{smar}{Y} + \textcolor{smar}{X} - 2\textcolor{smar}{W} - 2\textcolor{smar}{V} + 2\textcolor{smar}{U}),
    \end{aligned}
\end{equation}

\noindent where all terms are defined above.
\\

%%%%%%%%%%%%%%%%%%%%%%%%%%%%%%%%%%%%%%%%%%%%%%%%%%

% Don't change these lines
\bsp	% typesetting comment
\label{lastpage}
\end{document}